\documentclass[11pt]{article}
\usepackage{rotating}
\usepackage{epsfig}
\usepackage{latexsym}
\usepackage{graphicx}
\usepackage{psfrag}
\usepackage{amsmath,amssymb}
\def\ltsim{\lower3pt\hbox{$\, \buildrel < \over \sim \, $}}  
\def\gtsim{\lower3pt\hbox{$\, \buildrel > \over \sim \, $}}  
\makeatletter
\def\section{\@startsection {section}{1}{\z@}{-3.5ex plus -1ex minus
 -.2ex}{2.3ex plus .2ex}{\large\bf}}
\def\subsection{\@startsection{subsection}{2}{\z@}{-3.25ex plus -1ex
minus -.2ex}{1.5ex plus .2ex}{\normalsize\bf}}
\makeatother
\makeatletter
\def\theequation{\arabic{section}.\arabic{equation}}

\@addtoreset{equation}{section}
\renewcommand{\theequation}{\thesection.\arabic{equation}}
\makeatother

\newcommand{\captionfonts}{\small}
\makeatletter  
\long\def\@makecaption#1#2{%
  \vskip\abovecaptionskip
  \sbox\@tempboxa{{\captionfonts #1: #2}}%
  \ifdim \wd\@tempboxa >\hsize
    {\captionfonts #1: #2\par}
  \else
    \hbox to\hsize{\hfil\box\@tempboxa\hfil}%
  \fi
  \vskip\belowcaptionskip}
\makeatother   

\catcode`@=11
\def\marginnote#1{}
\newcount\hour
\newcount\minute
\newtoks\amorpm
\hour=\time\divide\hour
by60
\minute=\time{\multiply\hour by60 \global\advance\minute
by-\hour}
\edef\standardtime{{\ifnum\hour<12 \global\amorpm={am}
\else\global\amorpm={pm}\advance\hour by-12 \fi
 \ifnum\hour=0
\hour=12 \fi
 \number\hour:\ifnum\minute<10
0\fi\number\minute\the\amorpm}}
\edef\militarytime{\number\hour:\ifnum\minute<10
0\fi\number\minute}
\def\draftlabel#1{{\@bsphack\if@filesw
{\let\thepage\relax
 \xdef\@gtempa{\write\@auxout{\string
\newlabel{#1}{{\@currentlabel}{\thepage}}}}}\@gtempa
 \if@nobreak
\ifvmode\nobreak\fi\fi\fi\@esphack}
\gdef\@eqnlabel{#1}}
\def\@eqnlabel{}
\def\@vacuum{}
\def\draftmarginnote#1{\marginpar{\raggedright\scriptsize\tt#1}}
\def\draft{\oddsidemargin
0.0truein
 \def\@oddfoot{\sl preliminary draft \hfil
\rm\thepage\hfil\sl\today\quad\militarytime}
 \let\@evenfoot\@oddfoot
\overfullrule 3pt
 \let\label=\draftlabel
\let\marginnote=\draftmarginnote
\def\@eqnnum{(\theequation)\rlap{\kern\marginparsep\tt\@eqnlabel}
\global\let\@eqnlabel\@vacuum}
}
\catcode`@=12
\def\Xint#1{\mathchoice
   {\XXint\displaystyle\textstyle{#1}}%
   {\XXint\textstyle\scriptstyle{#1}}%
   {\XXint\scriptstyle\scriptscriptstyle{#1}}%
   {\XXint\scriptscriptstyle\scriptscriptstyle{#1}}%
   \!\int}
\def\XXint#1#2#3{{\setbox0=\hbox{$#1{#2#3}{\int}$}
     \vcenter{\hbox{$#2#3$}}\kern-.5\wd0}}

\def\dashint{\Xint-}

\def\bea{\begin{eqnarray}} \def\eea{\end{eqnarray}}
\def\be{\begin{eqnarray}} \def\ee{\end{eqnarray}} \def\nn{\nonumber}

\newcommand{\promille}{%
  \relax\ifmmode\promillezeichen
        \else\leavevmode\(\mathsurround=0pt\promillezeichen\)\fi}
\newcommand{\promillezeichen}{%
  \kern-.05em%
  \raise.5ex\hbox{\the\scriptfont0 0}%
  \kern-.15em/\kern-.15em%
  \lower.25ex\hbox{\the\scriptfont0 00}}

\begin{document}

\thispagestyle{empty}

\begin{center}
\hfill CERN-PH-TH/2007-129\\
\hfill IFT-UAM/CSIC-07-42\\
\hfill UAB-FT-633

\begin{center}

\vspace{1.7cm}

{\LARGE\bf Unparticles-Higgs Interplay}

\end{center}

\vspace{1.4cm}

{\bf Antonio Delgado$^{\,a}$, Jos\'e R. Espinosa$^{\,b}$ 
and Mariano Quir\'os$^{\,c}$}\\

\vspace{1.2cm}

${}^a\!\!$
{\em {Department of Physics, CERN-Theory division, Geneva 23, Switzerland}}

${}^b\!\!$
{\em { IFT-UAM/CSIC, Fac. Ciencias UAM, 28049 Madrid, Spain}}

${}^c\!\!$
{\em { IFAE, Universitat Aut{\`o}noma de Barcelona,
08193 Bellaterra, Barcelona (Spain)}}

{\em {and}}

{\em {Instituci\`o Catalana de Recerca i Estudis Avancats (ICREA)}}

\end{center}

\vspace{0.8cm}

\centerline{\bf Abstract}
\vspace{2 mm}
\begin{quote}\small
We show that scalar unparticles coupled to the Standard Model Higgs at
the renormalizable level can have a dramatic impact in the breaking of
the electroweak symmetry already at tree level. In particular one can
get the proper electroweak scale without the need of a Higgs mass term
in the Lagrangian. By studying the mixed unparticle-Higgs propagator
and spectral function we also show how unparticles can shift the Higgs
mass away from its Standard Model value, $2\lambda v^2$, and influence
other Higgs boson properties. Conversely, we study in some detail how
electroweak symmetry breaking affects the unparticle sector by
breaking its conformal symmetry and generating a mass gap. We also
show that, for Higgs masses above that gap, unparticles can increase
quite significantly the Higgs width.
\end{quote}

\vfill

\newpage
\section{Introduction}
In two recent papers~\cite{Georgi}, Georgi has proposed to look
seriously at the possibility that a conformal sector with a
non-trivial fixed point might be realized in nature and couple to our
standard world of particles.  He has shown how such sector would have
very unconventional features and, at least in the appropriate energy
range, will behave unlike a common particle sector. These two seminal
papers have been followed by a deluge of
work~\cite{Cheung:2007ue,multiunp} in all sorts of phenomenological
implications that such an unparticle sector could have.

In this paper we consider how unparticles could affect one of the
central issues of contemporary particle physics: the breaking of the
electroweak symmetry and the nature of the Higgs sector. After a brief
reminder of some aspects of unparticles relevant for this discussion
we show in section~2 how unparticles, if coupled to the Higgs operator
$|H|^2$ as recently considered in~\cite{shirman}, can have a dramatic
impact on electroweak symmetry breaking already at tree-level. In
section 3 we study the mixed unparticle-Higgs propagator and spectral
function and show: {\bf i)} How unparticles can shift the Higgs mass
away from its Standard Model (SM) value, $2\lambda v^2$, and,
conversely; {\bf ii)} How electroweak symmetry breaking affects the
unparticle sector by breaking its conformal symmetry and generating a
mass gap. For Higgs masses above that gap we also show that
unparticles can also affect significantly the Higgs width.

We will consider the ultraviolet (UV) coupling of an operator of
dimension $d_{UV}$ in the unparticle sector to the SM dimension-two
operator $|H|^2$ as
\be
\mathcal L=-\frac{1}{\mathcal M_U^{d_{UV}-2}}|H|^2\mathcal O_{UV}\ ,
\label{UVcoupling}
\ee
which flows in the infrared (IR) to
\be \mathcal L=-C_U \left(\frac{\Lambda_U}{\mathcal
    M_U}\right)^{d_{UV}-2}\Lambda_U^{2-d_U}|H|^2\mathcal O_{U}\equiv -\kappa_U
|H|^2\mathcal O_{U}\ ,
\label{IRcoupling}
\ee
where $d_U$ is the scaling dimension of the unparticle operator
$\mathcal O_{U}$ (usually considered in the interval $1<d_U<2$), $C_U$
is a dimensionless constant (whose value can be absorbed in the
definition of the scales $\Lambda_U$ and $\mathcal M_U$ and so it will
be fixed to one) and $\kappa_U$ has mass dimension $2-d_U$.

We take the tree-level Higgs potential 
\be
V_0=m^2 |H|^2+\lambda|H|^4\ ,
\label{tree}
\ee
where the squared mass parameter can have either sign or even vanish
and the quartic coupling $\lambda$ is related in the SM to the Higgs
mass at tree level by $m_{h0}^2=2\lambda v^2$ (for $m^2<0$). We write
the Higgs real direction as $Re(H^0)=(h^0+v)/\sqrt{2}$, with $v =246$
GeV.

The unparticle operator $\mathcal O_U$ coupled to $|H|^2$ in
Eq.~(\ref{IRcoupling}) has spin zero and its propagator
is~\cite{Georgi,Cheung:2007ue}
\be
P_U(p^2)=\frac{A_{d_U}}{2\sin(\pi d_U)} 
\frac{i}{(-p^2-i\epsilon)^{2-d_U}},\quad
A_{d_U}\equiv
\frac{16\pi^{5/2}}{(2\pi)^{2d_U}}\frac{\Gamma(d_U+1/2)}
{\Gamma(d_U-1)\Gamma(2d_U)}\ .
\label{prop}
\ee
The spectral function representation for this propagator
\be
\label{specrep}
-iP_U(p^2)=\int_0^\infty\frac{\rho_U(s)}{p^2-s+i\epsilon}\ ds\ ,
\ee
gives
\be
\label{espconf}
\rho_U(s)=\frac{A_{d_U}}{2\pi}s^{d_U-2}\ ,
\ee
with no poles and an essential singularity at $s=0$.

\section{Electroweak Breaking}
We are interested in the possible effects of the unparticle sector on
the Higgs sector through the coupling (\ref{IRcoupling}) and, in
particular, in examining the possible impact of unparticle effects on
electroweak symmetry breaking, in the spirit of~\cite{EQ}, which
analyzed this issue for standard hidden sectors.

The first observation, to which this paper is devoted, is that
important effects of the unparticle sector on the Higgs physics
already appear at tree level. When the Higgs field develops a non zero
vacuum expectation value (VEV) the conformal symmetry of the
unparticle sector is broken~\cite{shirman}.  From (\ref{IRcoupling})
we immediately see that in this non-zero Higgs background the physical
Higgs field will mix with the unparticle operator ${\cal O}_U$ and
moreover, a tadpole will appear for the operator ${\cal O}_U$ itself
which will therefore develop a non-zero VEV also.

In order to study these issues it is convenient to use a deconstructed
version of the unparticle sector, as discussed in~\cite{deco}. One
considers an infinite tower of scalars $\varphi_n$,
($n=1,...,\infty$), with masses squared $M_n^2=\Delta^2 n$. The mass
parameter $\Delta$ is small and eventually taken to zero, limit in
which one recovers a (conformal) continuous mass spectrum. It can be
shown~\cite{deco} that the deconstructed form of the operator ${\cal
O}_U$ is
\be
\label{OUdec}
{\cal O}\equiv \sum_n F_n \varphi_n \ ,
\ee
where 
\be
\label{Fdec}
F_n^2 = \frac{A_{d_U}}{2\pi}\Delta^2 (M_n^2)^{d_U-2}\ , \ee so that
the two-point correlator of ${\cal O}$ matches that of ${\cal O}_U$ in
the $\Delta\to 0$ limit. In the deconstructed theory then, the
unparticle scalar potential, including the coupling (\ref{IRcoupling})
to the Higgs field, reads
\be
\label{potdec}
\delta{V} = \frac{1}{2}\sum_n M_n^2\varphi_n^2+\kappa_U |H|^2\sum_n F_n 
\varphi_n\ .
\ee
A non-zero VEV, $\langle |H|^2\rangle=v^2/2$, would trigger then a VEV
for the fields $\varphi_n$:
\be
\label{vn}
v_n\equiv\langle\varphi_n\rangle=-\frac{\kappa_U v^2}{2M_n^2}F_n\ ,
\ee
thus implying
\be
\label{vevO}
\langle {\cal O}\rangle =\left\langle \sum_n F_n 
\varphi_n\right\rangle=-\frac{\kappa_U 
v^2}{2}\sum_n\frac{F_n^2}{M_n^2}\ .
\ee
In the continuum limit this gives
\be
\label{vevOU}
\langle {\cal O}_U \rangle = -\frac{\kappa_U
v^2}{2} \int_0^\infty \frac{F^2(M^2)}{M^2}dM^2\ ,
\ee
where 
\be
\label{F}
F^2(M^2)=\frac{A_{d_U}}{2\pi}(M^2)^{d_U-2}\ ,
\ee
is the continuum equivalent of (\ref{Fdec}). We immediately see that
$\langle {\cal O}_U \rangle$ has an IR divergence. This is due to the
fact that for $M\to 0$ the tadpole diverges while the mass itself,
that should stabilize the unparticle VEV, goes to zero.

As a possible cure for this divergence problem one can envisage
several possibilities. One might try to introduce quartic couplings
$(1/4)\lambda_n \varphi_n^4$. A finite non-zero continuum limit
requires that $\lambda_n$ scales with $\Delta$ as $\lambda_n\sim
\mu_\lambda^2/\Delta^2$, where $\mu_\lambda$ is some mass
parameter. Scale invariance requires in fact that
$\mu^2_\lambda\propto M^2$ and this again does not solve the IR
problem of $\langle {\cal O}_U\rangle$. Other alternatives, like
introducing an ${\cal O}_U^2$ term, also fail in this respect. In this
paper we consider instead introducing an IR-regulator related to the
breaking of the conformal symmetry by the Higgs VEV. We show below
that this indeed stabilizes $\langle {\cal O}_U\rangle$.

One can easily get an IR regulator in (\ref{F}) by including a
coupling~\footnote{Notice that this coupling cannot be expressed in
terms of ${\cal O}_U$ in the continuum limit.}
\be
\label{HHUU}
\delta V = \zeta |H|^2 \sum_n \varphi_n^2\ ,
\ee
in the deconstructed theory. This coupling respects the conformal
symmetry but will break it when $H$ takes a VEV. Now one gets
\be
\label{vnIR}
v_n\equiv\langle\varphi_n\rangle=-\frac{\kappa_U v^2}{2(M_n^2+\zeta 
v^2)}\ F_n\ ,
\ee
leading in the continuum limit to
\be u(M^2)\equiv -\frac{\kappa_U v^2}{2}
\frac{F(M^2)}{M^2+\zeta v^2}\ ,
\label{u}
\ee
[where $u(M^2)$ is the continuum version of the unparticle VEV, scaled
as $v_n=\Delta u_n$], and then to
\be
\langle {\cal O}_U \rangle = -\frac{\kappa_U v^2}{2} \int_0^\infty
\frac{F^2(M^2)}{M^2+\zeta v^2}\ dM^2\ .
\ee
This integral is obviously finite for $1<d_U<2$ and yields explicitly
\be
\langle {\cal O}_U \rangle
=-\frac{1}{2}\kappa_U\frac{A_{d_U}}{2\pi}\zeta^{d_U-2}
v^{2d_U-2}\Gamma(d_U-1)\Gamma(2-d_U)\ .
\label{vevOUIR}
\ee

In the presence of (\ref{HHUU}) the minimization condition for the
Higgs VEV $v$ is then
\be
m^2 +\lambda v^2+\kappa_U \sum_n F_n v_n+\zeta \sum_n v_n^2=0\ ,
\label{mincaso1}
\ee
or, in the continuum limit, 
\be m^2 +\lambda v^2 + \int_0^\infty dM^2\left[\kappa_U
  F(M^2)+\zeta u(M^2)\right]u(M^2)=0\ ,
\label{min1}
\ee
which, using the VEV (\ref{u}), translates into
\be
m^2+\lambda v^2 -\lambda_U (\mu_U^2)^{2-d_U}v^{2(d_U-1)}=0\ ,
\label{minimoew}
\ee
with
\be
\label{lambdaU}
\lambda_U\equiv \frac{d_U}{4} \zeta^{d_U-2}\Gamma(d_U-1)\Gamma(2-d_U)\ ,
\ee
and
\be
(\mu_U^2)^{2-d_U}\equiv \kappa_U^2 \frac{A_{d_U}}{2\pi}\ .
\ee
We see that the effect of the unparticles in the minimization equation
(\ref{minimoew}) is akin to having a Higgs term $h^{2d_U}$ in the
potential, that is, for $1<d_U<2$, a term somewhere between $h^2$ and
$h^4$!  Notice also that condition (\ref{minimoew}) can be easily
satisfied since the term induced by the unparticle VEV is negative. In
particular, for $m^2=0$ the Higgs VEV is induced by its coupling to
unparticles as
\be 
v^2=\left(\frac{\lambda_U}{\lambda}\right)^{\frac{1}{(2-d_U)}}\mu_U^2\ ,
\label{VEV}
\ee
and it is therefore determined by the mass parameter $\mu_U$.
In terms of the fundamental scales $\Lambda_U$ and $\mathcal M_U$ in 
(\ref{IRcoupling}) this mass parameter reads
\be 
\mu_U^2\equiv \left(\frac{A_{d_U}}{2\pi}\right)^{\frac{1}{2-d_U}} 
\left(\frac{\Lambda_U^2}{\mathcal M_U^2}\right)^{\frac{d_{UV}-2}{2-d_U}}\,
 \Lambda_U^2 \ ,
\label{muU2}
\ee
and one can easily get $\mu_U\sim v$ from the scales $\Lambda_U\gg v$
and $\mathcal M_U\gg \Lambda_U$ provided that $d_{UV}>2$. For later
numerical work we will usually take $\kappa_U=v^{2-d_U}$, which
corresponds to $\mu_U^2=\mu_v^2\equiv
v^{2}[A_{d_U}/(2\pi)]^{1/(2-d_U)}$.

\begin{figure}[htb]
\psfrag{dU}[][bl]{$d_U$}\psfrag{la}[][l]{$\lambda$}
\hspace{5cm}
\begin{center}
\epsfig{file=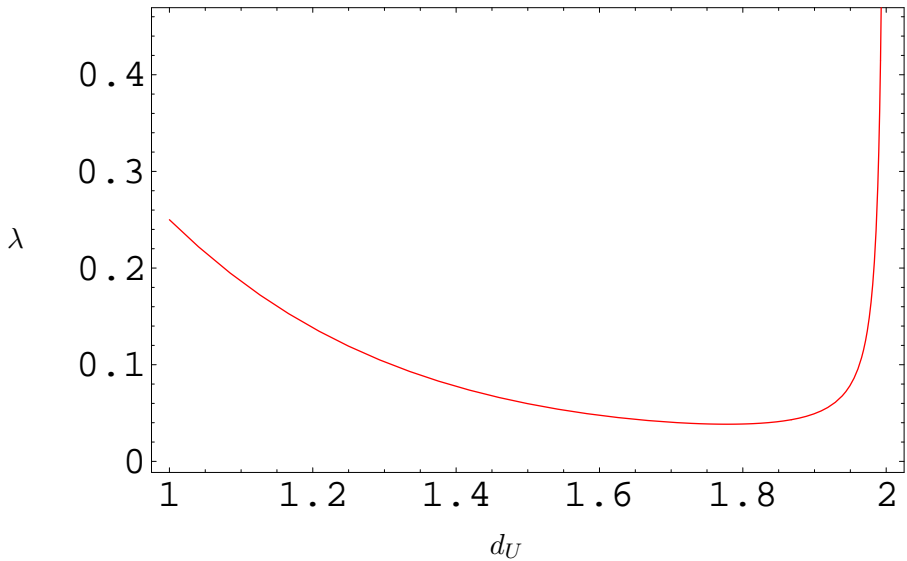,width=0.8\textwidth}
\end{center} 
\caption{\it Plot of $\lambda$ from the
  minimization condition (\ref{fixing}) for the case $m=0$, $\zeta=1$ and
$\mu_U^2=\mu_v^2$ as a
  function of $d_U$.  }
\label{lambda1}
\end{figure}

Electroweak symmetry breaking at tree level requires the condition
\be
m^2\leq \lambda_U (\mu_U^2)^{2-d_U}v^{2(d_U-1)}\ ,
\label{ewb1}
\ee
in which case the Higgs potential has a Mexican-hat shape. In the
particular case of $m^2=0$, condition (\ref{ewb1}) is automatically
satisfied.  Of course one has to adjust the parameters in
(\ref{minimoew}) to have the minimum at the correct value. This 
requires that the Higgs quartic coupling is chosen as
\be
\lambda=-\frac{m^2}{v^2}+\lambda_U (\mu_U^2)^{2-d_U}v^{2(d_U-2)}\ ,
\label{fixing}
\ee
which shows how unparticles modify the usual Standard Model relation.
A plot of $\lambda$ as a function of $d_U$ is shown in
Fig.~\ref{lambda1} for the case $m=0$, $\mu_U^2=\mu_v^2$ and
$\zeta=1$.  The scaling of $\lambda$ with $\mu_U$ and $\zeta$ can be
read off from Eq.~(\ref{fixing}).

\section{Pole Mass and Spectral Function}
Having found a way of stabilizing the unparticle (and Higgs) VEVs
keeping $\langle {\cal O}_U \rangle$ finite we can move on to the
study of the combined Higgs-unparticle propagator. Perhaps the
simplest way to obtain this propagator is to start with the
deconstructed theory.  The neutral component of the Higgs, $h^0$,
mixes with the $\varphi_n$ fields in an infinite scalar mass matrix,
but the secular equation can easily be obtained. Taking its continuum
limit one obtains the corresponding propagator for the coupled
Higgs-unparticle system (that re-sums unparticle corrections):
\be
iP(p^2)^{-1}= p^2 - m_{h0}^2 + v^2(\mu_U^2)^{2-d_U} \int_0^\infty
\frac{(M^2)^{d_U-2}}{M_U^2(M^2)-p^2} r(M^2)dM^2\ , 
\ee 
where $M_U^2(M^2)$ is the
mass distribution of unparticles after conformal symmetry breaking:
\be
M_U^2(M^2)=M^2+\zeta v^2\ ,
\; \; \; \; \;
{\rm and}
\; \; \; \; \; 
 r(M^2)=\left(\frac{M^2}{M^2+\zeta v^2} \right)^2\ .
\ee
In order to understand the interplay between the Higgs and the
unparticle sector after electroweak symmetry breaking it is
instructive to examine the spectral representation of this
propagator, which can be obtained easily.  

There are two qualitatively different cases, depending on whether the
Higgs mass squared $m_h^2$ is larger or smaller than
$m_{gap}^2\equiv\zeta v^2$.  Here $m_h^2$ is the Higgs mass corrected
by the interactions to unparticles and implicitly given by the pole
equation
\be
\label{mhshift}
m_h^2=m_{h0}^2 - v^2(\mu^2_U)^{2-d_U} \int_0^\infty
\frac{(M^2)^{d_U-2}}{M_U^2(M^2)-m_h^2}\ r(M^2)dM^2\ .
\ee

{\bf 1.} Let us first consider the case $m_h^2<\zeta v^2$. The
analytical equation for $m_h^2$ can be explicitly written as
\bea
m_h^2&=&m_{h_0}^2- \frac{v^2(\mu^2_U)^{2-d_U}
}{m_h^4}\Gamma(d_U-1)\Gamma(2-d_U)\times\nn\\
&&\times
\left[(\zeta v^2-m_h^2)^{d_U}
+d_U m_h^2(\zeta v^2)^{d_U-1}-(\zeta v^2)^{d_U}\right]\ .
\label{pole1}
\eea
Notice that the last term in (\ref{pole1}) goes to zero in the
(particle) limit $d_U\to 1$ and therefore in this limit the pole mass
is the standard one, $m_h^2=m_{h_0}^2$.

\begin{figure}[htb]
\psfrag{mh}[][bl]{$m_h$}\psfrag{dU}[][l]{$d_U$}
\begin{center}
\epsfig{file=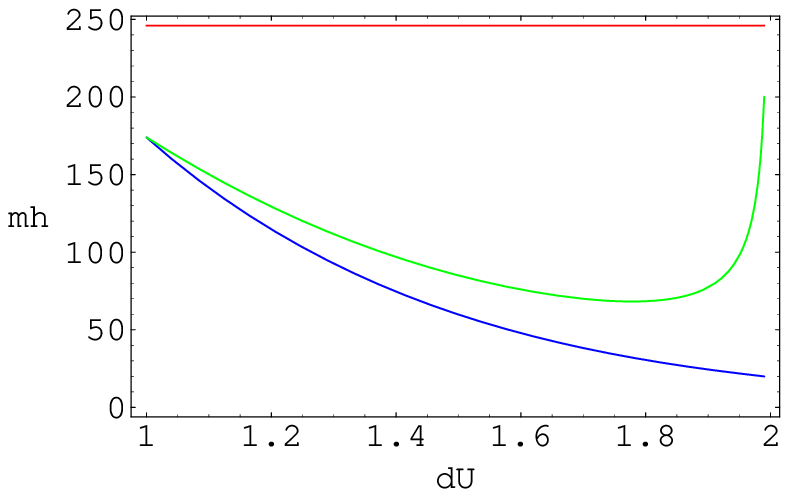,width=0.8\textwidth}
\end{center} 
\caption{\it Plot of the pole Higgs mass $m_h$ (lower curve) and
unresummed Higgs mass $m_{h_0}$ (upper curve) as functions of $d_U$
for $\mu_U^2=\mu_v^2$ and $\zeta=1$. The straight 
line is $m_{gap}$. Masses are in GeV.}
\label{mh1}
\end{figure}

The spectral function is explicitly given by
\be
\label{rho1}
\rho(s)= \frac{1}{K^2(m_h^2)} \delta(s-m_h^2)+\theta(s-\zeta v^2) 
\frac{Q^2_{U}(s)}{\mathcal{D}^2(s)+\pi^2Q_{U}^4(s)}\ ,
\ee
with
\be
\label{QU}
Q^2_U(s)\equiv v^2(\mu_U^2)^{2-d_U}
\frac{(s-\zeta v^2)^{d_U}}{s^2}\ ,
\ee
and
\be
\label{Ds}
\mathcal{D}(s)\equiv {\mathrm P.V.}\left[iP(s)^{-1}\right]=s -
m_{h0}^2 + v^2(\mu_U^2)^{2-d_U} \dashint_0^\infty
\frac{(M^2)^{d_U-2}}{M_U^2(M^2)-s} r(M^2)dM^2\ , \ee
where the slash in the integral denotes that its principal value
should be taken. An explicit expression for $\mathcal{D}(s)$ can also
be obtained analytically from (\ref{pole1}). Finally,
\be
\label{Kmh}
K^2(m_h^2)\equiv 
\left.\frac{d}{ds}\mathcal{D}(s)\right|_{s=m_h^2}\ ,
\ee
which in this case reads
\be
K^2(m_h^2)= 1+v^2(\mu_U^2)^{2-d_U}\int_0^\infty 
\frac{(M^2)^{d_U-2}}{\left[M_U^2(M^2)-m_h^2\right]^2}r(M^2)dM^2\ .
\ee

We first notice from Eq.~(\ref{pole1}) that the Higgs mass at tree
level is no longer simply given by $m_{h0}^2$ but it is shifted by a
negative amount by the effect of the coupling to
unparticles~\footnote{It is easy to prove that the function in the
square brackets in (\ref{pole1}) is positive definite for
$m_h^2<m_{gap}^2$.}. In Fig.~\ref{mh1} we plot the pole mass $m_h$ as
a function of $d_U$ for $\mu_U^2=\mu_v^2$, $\zeta=1$ and compare it
with $m_{h_0}$. In this case we observe that $m_h^2<m_{gap}^2$ for all
values of $d_U$.
\begin{figure}[htb]
\psfrag{rho}[][bl]{$\zeta v^2\,\rho$}\psfrag{s}[][l]{$s/\zeta v^2$}
\begin{center}
\epsfig{file=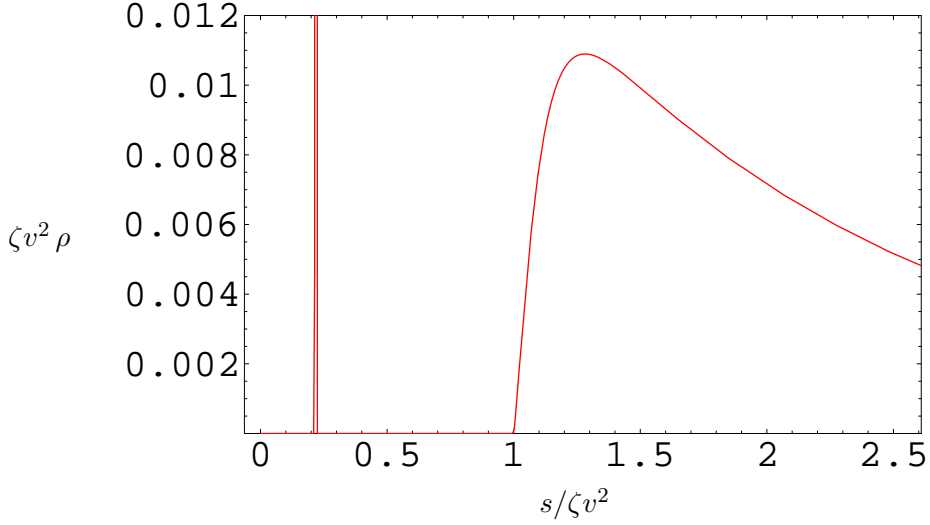,width=0.8\textwidth}
\end{center} 
\caption{\it Spectral function $\rho$ as a function of $s$ for
$\mu_U^2=\mu_v^2$, $\zeta=1$ and $d_U=1.2$. All dimensions are scaled
with $\zeta v^2$.}
\label{fig:rho1}
\end{figure}
On the other hand, the coupling of the unparticle sector to the Higgs
sector, that breaks the conformal symmetry, results in a modification
of the ``unparticle part'' of the spectral function [the second term
in (\ref{rho1})]. It still has no poles but now there is a mass gap,
$m_{gap}$. The shape of the spectral function (\ref{rho1}) is shown in
Fig.~\ref{fig:rho1}, where we have chosen $\mu_U^2=\mu_v^2$, $\zeta=1$
and $d_U=1.2$, and the Higgs masses obtained from Fig.~\ref{mh1} are
$m_h=115$ GeV and $m_{h_0}=130$ GeV. All dimensional quantities are
made dimensionless by scaling them with $\zeta v^2$. This result for
the spectral function has some similarities with that introduced in
Refs.~\cite{shirman,terning} although it has been obtained through a
different approach and differs from theirs.

Due to this mixing with the unparticles, the Higgs properties will
also be affected in a way similar to the usual singlet
mixing~\cite{bij}. It is straightforward to obtain that the
Higgs-composition of the isolated resonance at $m_h$, call it $R_h$,
is simply
\be
\label{Rh}
R_h=\frac{1}{K(m_h)}\ ,
\ee
where $R_h=1$ would correspond to a pure SM Higgs with no unparticle
admixture. Conversely, the unparticle continuum gets the
Higgs-composition that the Higgs has lost, distributed through the
$M^2$-dependent function
\be
\label{Ru}
R_U(M^2)=-\theta(M^2-m_{gap}^2)
\frac{Q_U(M^2)}{(M^2-m_h^2)K(m_h^2)}\ .
\ee
Note that, unlike $R_h$, the quantity $R_U(M^2)$ is a Higgs-component
density and therefore has mass dimension -1.  One can check that the
following sum rule \be R_h^2 + \int_0^\infty R_U^2(M^2)\ dM^2 = 1\ ,
\ee holds. The quantities $R_h$ and $R_U(M^2)$ play a major role in
the phenomenology of Higgs and unparticles after electroweak symmetry
breaking.

\begin{figure}[htb]
\psfrag{mh}[][bl]{$m_h$}\psfrag{dU}[][l]{$d_U$}
\begin{center}
\epsfig{file=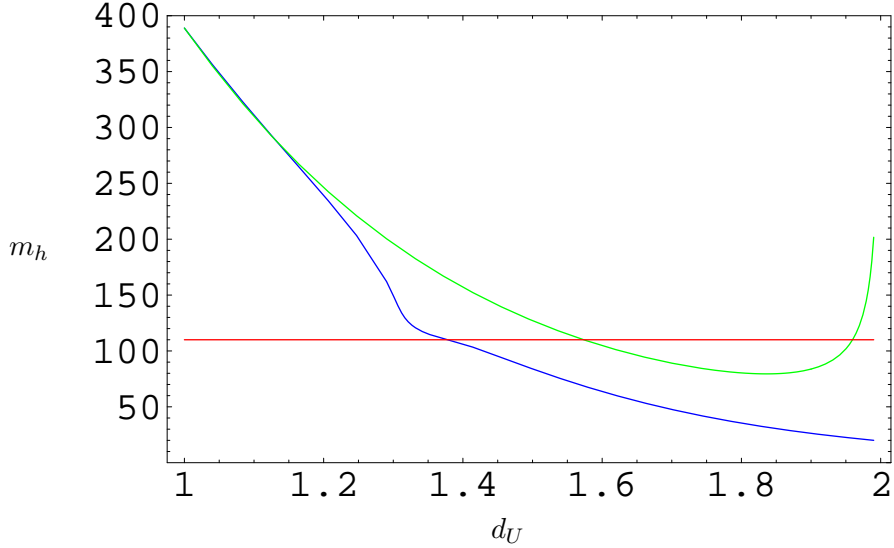,width=0.8\textwidth}
\end{center} 
\caption{\it Plot of the pole Higgs mass $m_h$ (lower curve) and
unresummed Higgs mass $m_{h_0}$ (upper curve) as functions of $d_U$
for $\mu^2_U=\mu^2_v$ and $\zeta=0.2$. The straight line is
$m_{gap}$. Masses are in GeV.}
\label{mh1prime}
\end{figure}

\begin{figure}[htb]
\psfrag{mh2}[][bl]{$m_h/m_{gap}$}\psfrag{dU}[][l]{$d_U$}
\begin{center}
\epsfig{file=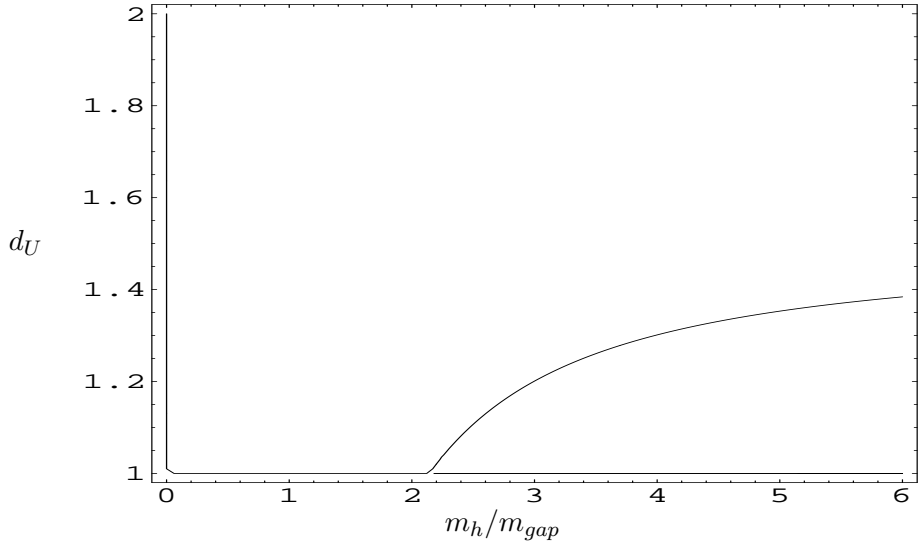,width=0.8\textwidth,height=0.5\textwidth}
\end{center} 
\caption{\it Plot of the sign of the shift in the pole Higgs mass $m_h^2$ 
with respect to the SM value $m^2_{h0}=2\lambda v^2$ as a function 
of $m_h/m_{gap}$ and $d_U$. This shift is negative above the line shown 
and positive below it.}
\label{fig:masshift}
\end{figure}

{\bf 2.} If $m_h^2>m^2_{gap}$, the delta function for the Higgs pole
merges with the unparticle continuum. Before showing this explicitly,
we first notice that the integrand in (\ref{mhshift}) crosses a pole
and the principal value of the integral should be taken. This feature
is exhibited in Fig.~\ref{mh1prime} where we plot the pole mass $m_h$
as a function of $d_U$, for $\mu_U^2=\mu_v^2$ and $\zeta=0.2$, and
compare it with $m_{gap}$. We see that in the region $d_U\ltsim 1.4$
($d_U\gtsim 1.4$) $m_h^2\gtsim m^2_{gap}$ ($m_h^2\ltsim
m^2_{gap}$). At the value $d_U\simeq 1.4$ there is a kink in the
integral (\ref{mhshift}) because the principal value has been
taken. The analytical equation for $m_h^2$ now reads:
\bea
m_h^2&=&m_{h_0}^2- \frac{v^2(\mu^2_U)^{2-d_U}
}{m_h^4}\Gamma(d_U-1)\Gamma(2-d_U)\times\nn\\
&&\times
\left[(m_h^2-\zeta v^2)^{d_U}\cos(\pi d_U)
+d_U m_h^2(\zeta v^2)^{d_U-1}-(\zeta v^2)^{d_U}\right]\ .
\label{pole2}
\eea
One can also show that it is possible to have a positive shift in the
Higgs mass, getting $m_h>m_{h0}$, for sufficiently large $m_h/m_{gap}$
and small enough $d_U$. The sign of the Higgs mass shift is shown in
Fig.~\ref{fig:masshift} where the positive sign corresponds to the
region connected with the lower right corner.
\begin{figure}[htb]
\psfrag{rho}[][bl]{$\zeta v^2\,\rho$}\psfrag{s}[][l]{$s/\zeta v^2$}
\begin{center}
\epsfig{file=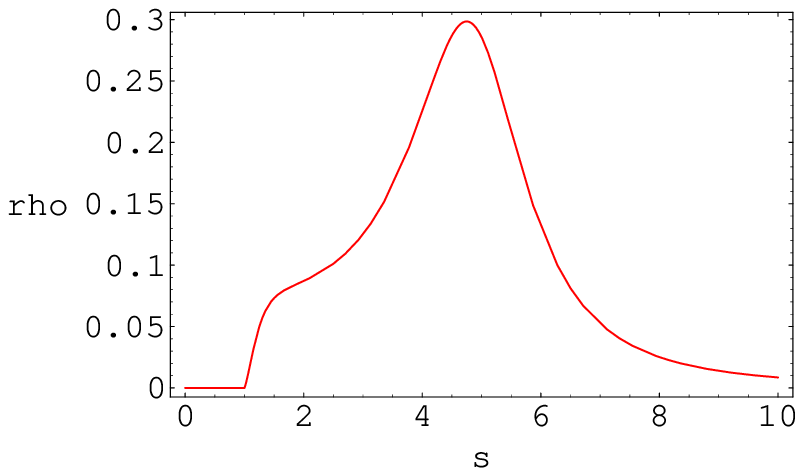,width=.8\textwidth}
\end{center} 
\caption{\it Spectral function $\rho$ as a function of $s$ for
$\mu^2_U=\mu_v^2$, $\zeta=0.2$ and $d_U=1.2$. All dimensions are
scaled with $\zeta v^2$.}
\label{rho1prime}
\end{figure}
The spectral function in this case simply reads
\be
\label{rho1b}
\rho(s)=\theta(s-\zeta v^2)
\frac{Q^2_{U}(s)}{\mathcal{D}^2(s)+
\pi^2Q_{U}^4(s)}\ ,
\ee
with $Q^2_{U}(s)$ as given in (\ref{QU}). Near the Higgs pole one can 
approximate 
\be
\label{nearmh}
\mathcal{D}(s)\simeq (s-m_h^2)K^2(m_h^2)\ ,
\ee
where $K^2(m_h^2)$ is defined in Eq.~(\ref{Kmh}).  In this case one
should be careful about using the principal value definition of
$\mathcal{D}(s)$ to calculate properly its derivative at $m_h^2$. In
fact an analytical expression for $K^2(m^2_h)$ in this case can be
simply obtained by taking the derivative of (\ref{pole2}).

The shape of this spectral function is shown in Fig.~\ref{rho1prime}
where we have chosen $\mu^2_U=\mu_v^2$, $\zeta=0.2$ and $d_U=1.2$, and
the Higgs masses obtained from Fig.~\ref{mh1prime} are $m_h=240$ GeV
and $m_{h_0}=245$ GeV.  The peak in Fig.~\ref{rho1prime} is due to the
merging of the Higgs with unparticles.

\begin{figure}[htb]
\psfrag{Gamma}[][bl]{$\Gamma_h$/GeV}\psfrag{dU}[][l]{$d_U$}
\begin{center}
\epsfig{file=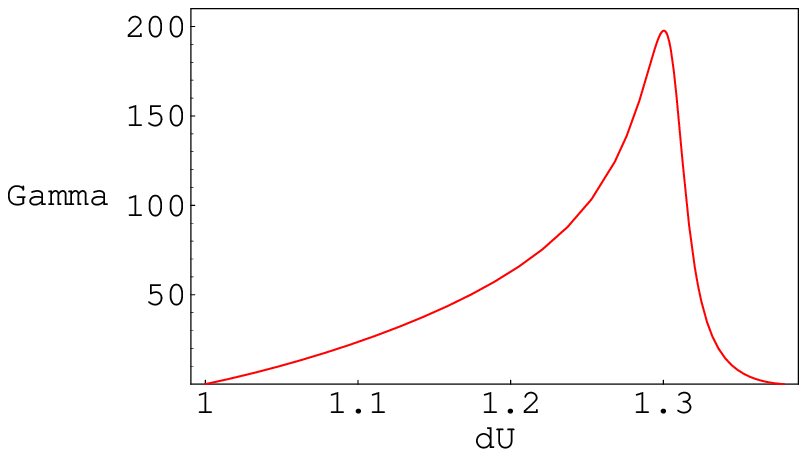,width=0.8\textwidth}
\end{center} 
\caption{\it Width of the Higgs boson from unparticle merging as a function of
$d_U$ for $\mu^2_U=\mu_v^2$ and $\zeta=0.2$}
\label{width1}
\end{figure}

Inserting (\ref{nearmh}) in the spectral function (\ref{rho1b}) we see
that the Higgs resonance has a Breit-Wigner shape of width
\be
\label{width}
\Gamma_h = \theta(m_h^2-m_{gap}^2)\frac{\pi 
Q^2_{U}(m_h^2)}{m_h K^2(m_h^2)}\ .
\ee
This width $\Gamma_h$ can be extremely wide ($\sim$ 100 GeV) depending
on the parameter choices and it is plotted as a function of $d_U$ in
Fig.~\ref{width1}. We can see from Fig.~\ref{width1} that (as
expected) it is different from zero only in the region where
$m_h>m_{gap}$. Needless to say this kind of effect can dramatically
modify the expectations for Higgs searches.

\section{Conclusions}

In this paper we have investigated the possibility of coupling the
Higgs boson to a conformal sector of unparticles, of the type recently
proposed by Georgi. A first consequence of that coupling is that
electroweak symmetry breaking generates a tadpole for the unparticles.
That tadpole would destabilize the theory in the absence of new
interaction terms that keep the unparticle VEV finite. We have
introduced for that purpose a new interaction between the Higgs and
the unparticle sector using a deconstructed version of the latter.

Having stabilized the unparticle VEV we have a consistent framework in
which to study the mutual influence between the Higgs and the
unparticle sectors. We find changes in the properties of the Higgs
(like its mass and its width) already at tree level, making the Higgs
and the unparticles a mixed sector. Studying the propagator and the
spectral function of this sector we find that there is a single pole,
corresponding to the Higgs, with the pole mass no longer given just by
the SM value, $2\lambda v^2$.  We also find a mass gap in the formerly
continuous spectral function for the unparticles, clearly indicating
that the conformal symmetry has been broken. This was expected from
previous work in the literature but we are able to discuss this
breaking explicitly.

When the Higgs mass is greater than the unparticle mass gap, the Higgs
can decay into unparticles and acquires a width which can be, in
principle, very large. This can have dramatic consequences for Higgs
searches at the LHC since it will mean that the Higgs will decay
invisibly unless these unparticles are also coupled to the SM sector
and have a sufficiently short decay length. As a last comment we can
say that this is another example of how the Higgs can be the window to
new sectors which would be completely hidden to us otherwise.

\subsection*{Acknowledgments}

\noindent 
Work supported in part by the European Commission under the European
Union through the Marie Curie Research and Training Networks ``Quest
for Unification" (MRTN-CT-2004-503369) and ``UniverseNet"
(MRTN-CT-2006-035863); by a Comunidad de Madrid project (P-ESP-00346);
and by CICYT, Spain, under contracts FPA 2004-02012, FPA 2004-02015
and FPA 2005-02211.

\end{document}